\documentclass[review]{elsarticle}

\usepackage{geometry}
\journal{Elsevier}

\biboptions{authoryear}

\date{}
\usepackage{booktabs}
\usepackage{siunitx}
\usepackage{amsmath}
\usepackage{subfigure}
\usepackage{float}
\usepackage[capitalize]{cleveref}
\usepackage{xcolor}
\usepackage{amsfonts}
\usepackage{bm}
\usepackage{diagbox}
\usepackage{url} 
\usepackage{algpseudocode}
\usepackage[linesnumbered,algoruled,boxed,lined]{algorithm2e}

\usepackage{caption}
\DeclareMathOperator*{\argmax}{arg\,max}

\begin{document}

\begin{frontmatter}

\title{Performance evaluation of an offshore wave measurement buoy in monochromatic waves}

\author[mymainaddress]{Xuepeng Fu\corref{mycorrespondingauthor}}
\cortext[mycorrespondingauthor]{Corresponding author}
\ead{xuepeng.fu@nlr.gov}

\author[mymainaddress]{Frederick Driscoll}
\author[mymainaddress]{Rebecca Fao}
\author[mymainaddress]{Calum Kenny}
\author[mymainaddress]{Kevin Patrick Griffin}
\author[mymainaddress]{Mark Murphy}
\author[mymainaddress]{Scott Lambert}
\address[mymainaddress]{National Laboratory of the Rockies, Golden, CO 80401, USA}

\begin{abstract}
The accurate measurement of waves underpins marine energy resource characterization, device design, and project development. Datawell wave buoys are widely deployed around the world and have long served as a trusted standard for wave measurements. We quantify the measurement performance, including wave elevation and energy flux estimation, of a Datawell DWR-MkIII buoy using prescribed monochromatic heave motions on a large-amplitude six-degree-of-freedom motion platform at the National Laboratory of the Rockies, assuming the buoy behaves as an ideal wave follower. Commanded motions were validated with an optical motion tracking system while buoy elevation and raw acceleration were recorded. Wave elevations were propagated to wave energy flux estimation using four methods, including one frequency-domain method and three time-domain methods. The Bayesian optimization was applied for design of experiments, and records from three test sites were also applied and evaluated in the present study. Results show two error regions within the nominal period range of $\SI{1.6}{s}$ to $\SI{30}{s}$. For wave periods between $\SI{5}{s}$ and $\SI{25}{s}$, the buoy provides accurate wave height measurements. For short periods less than $\SI{5}{s}$, the $\SI{1.28}{Hz}$ sampling frequency induces sub-Nyquist artifacts that bias elevation and can drive maximum energy flux estimation errors above $100\%$. For long periods exceeding $\SI{25}{s}$, the buoy reported elevation is underpredicted with error depending on period but relatively independent of wave height, with maximum wave height and wave energy flux errors reaching $64\%$ and $87\%$, respectively. The analysis of field data also indicates that the currently recommended method for estimating wave energy flux may underestimate the wave energy flux. 
\end{abstract}


\end{frontmatter}


\section{Introduction}
Ocean wave energy is a renewable resource that is predictable up to several days in advance, available continuously throughout the day, and characterized by a relatively high energy density compared with other renewable sources \citep{lehmann2017ocean}. In the United States, wave energy accounts for approximately $83\%$ of the total ocean hydrokinetic energy resource \citep{nrel2021renewable}. Wave resource characterization is the first step in marine energy development, as it directly informs wave energy converter (WEC) design and deployment \citep{ahn2020wave}. First, it specifies the environmental load inputs used throughout the WEC design process, consistent with established ocean engineering practice \citep{chen2021calibration,faltinsen1993sea}. Second, accurate estimation of wave energy flux is critical for assessing WEC performance and efficiency. Therefore, effective utilization of this resource depends on accurate characterization of wave energy flux in both coastal and offshore environments \citep{ahn2020wave}.

Two main approaches are used to characterize the wave energy resource: numerical modeling and direct wave measurements. The first method employs numerical wave models, such as SWAN \citep{booij1996swan} or WAVEWATCH III \citep{tolman2009user}, to simulate wave propagation and estimate wave energy flux across broad regions and long times. Researchers have generated multi-decade wave hindcasts using such models \citep{arinaga2012atlas,gonccalves201833}, and a scalable methodology for wave resource assessment has been developed to address previous limitations and improve the consistency of wave resource estimates \citep{kilcher2023scalable}. The second method relies on offshore wave buoys, which directly record wave height, period, and spectra at specific sites \citep{krogstad1999some,herbers2012observing,pillai2021framework}. Buoy observations provide high-quality data for calculating local wave energy flux and are crucial for validating and calibrating numerical model predictions \citep{hasselmann1973measurements}. In practice, these approaches are complementary: buoy data validate numerical models while model simulations fill spatial and temporal gaps between buoy locations, yielding a more accurate and comprehensive wave resource characterization.

Ensuring the accuracy and consistent application of these methodologies is essential to the marine energy community. The International Electrotechnical Commission (IEC) Technical Committee on Marine Energy has thus issued technical specifications governing site-focused wave resource assessment, notably, IEC~62600-100 \citep{IEC62600-100-2024} and IEC~62600-101 \citep{IEC62600-101-2024}. These technical specifications provide internationally accepted guidance on model configuration and validation procedures and have been applied in studies worldwide \citep{ramos2016exploring,hemer2017revised,lokuliyana2020sri,garcia2021wave}.

Accelerometer-based offshore wave measurement buoys, such as the Datawell Directional Waverider MkIII (DWR-MkIII), are widely considered the industry standard for wave measurements and serve as indispensable ground truth in wave energy studies \citep{o1996comparison,andrews2019evaluation}. The IEC technical specifications require calibrated buoys, and the DWR-MkIII is among the few commercially available instruments that meet this requirement \citep{IEC62600-100-2024}. The Waverider series buoys have also been used as the standard for validating next-generation buoys such as the Spotter buoy \citep{de2003field,colbert2010field,liu2015performance,raghukumar2019performance,beckman2022quantifying,jangir2023comparative,jangir2023comparativetwo}. While previous studies have explored sources of measurement uncertainty, most have focused on external influences, such as mooring configuration and the hydrodynamic response associated with the mooring system \citep{draycott2022experimental} or biofouling effects that can dampen buoy motion and distort wave spectra over long-term deployments \citep{thomson2015biofouling}. 

\begin{figure}[ht!]
    \centering
    \includegraphics[width=.25\textwidth]{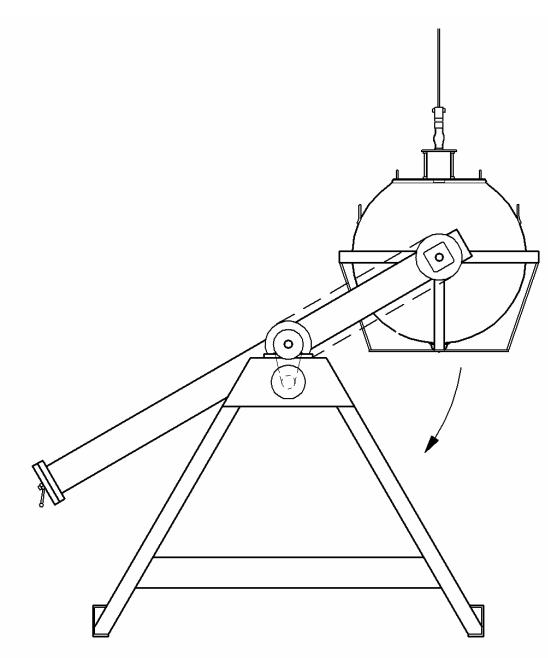}
    \caption{Ferris-wheel buoy calibration device. The buoy is fixed in the apparatus and forced to move through a circular trajectory, simulating its response.}
  \label{calidevice}
\end{figure}

Before ocean deployment, buoys are typically calibrated and verified using dry tests on a Ferris-wheel-like apparatus \citep{gerritzen1993calibration} as shown in \cref{calidevice}, in which the buoy is fixed and driven along a circular trajectory that is treated as an ideal wave follower. However, such Ferris-wheel setups can provide only a limited set of prescribed motions with fixed heave amplitude and thus do not fully characterize buoy performance over the broader range of motions relevant to field deployment, which is still used in the present buoy engineering practice. The marine energy field demands higher accuracy in wave measurements than many traditional ocean engineering applications, particularly for wave energy estimation. Therefore, buoy performance should be evaluated over a wider range of prescribed motions. As an early effort, \citet{van2018wave} evaluated the dynamic response of a DWR-MkIII buoy under prescribed motions using a motion-simulator platform at the Maritime Research Institute Netherlands, and reported potential errors in long-wave conditions, such as swell or bound second-order waves in shallow water \citep{ashton2015errors}. However, the study was limited to a narrow range of wave periods and did not examine potential energy flux estimation performance degradation at the bounds of the buoy operational range.

In the present study, we introduce the National Laboratory of the Rockies' (NLR's) dry test work for the intrinsic performance evaluation of the wave measurement buoy on the large-amplitude motion platform (LAMP). Compared with traditional calibration procedures, LAMP can generate a much wider range of prescribed motion cases than a Ferris-wheel apparatus, which can cover all the buoy operational period range \citep{Datawell2025WaveriderManual}, with heave amplitudes up to $\SI{0.9}{m}$ \citep{Friedman2025LAMP}. The performance evaluation is conducted based on DWR-MkIII buoy with the prescribed monochromatic heave motion experiment. The buoy is also assumed as a perfect wave follower here, and commanded motions were validated with the Qualisys optical system. Buoy elevation was recorded at \SI{1.28}{Hz} using the Wave5 software, and raw acceleration was recorded by the buoy at \SI{1.28}{Hz} using the Buoy Tester software to quantify the wave height measurement error. The test periods fall within the buoy’s operational limits. We propagated these errors to wave energy flux estimations using four methods: a frequency-domain method, a small-amplitude time-domain method, a Hilbert transform-based method, and a wavelet-based method. Finally, three Coastal Data Information Program (CDIP) field records were reprocessed to examine practical implications. This study presents an potential experimental approach for developing a new wave buoy calibration method and identifies key uncertainty sources affecting wave energy estimation, with the longer-term aim of informing future IEC technical specifications for marine energy development.

\section{Experimental setup}\label{expsec}
\subsection{Experimental apparatus}
The experiment was conducted on LAMP at the National Laboratory of the Rockies. This motion platform can act six degrees-of-freedom (6-DOFs) motion and is capable of translational strokes of up to $\pm1.25\ \text{m}$ in surge, $\pm1.15\ \text{m}$ in sway, and $\pm0.9\ \text{m}$ in heave, and rotational strokes of up to $\pm25.5^\circ$ in roll, $\pm30.0^\circ$ in pitch, and $\pm26^\circ$ in yaw \citep{Friedman2025LAMP}.

\begin{figure}[ht!]
    \centering    \includegraphics[width=.99\textwidth]{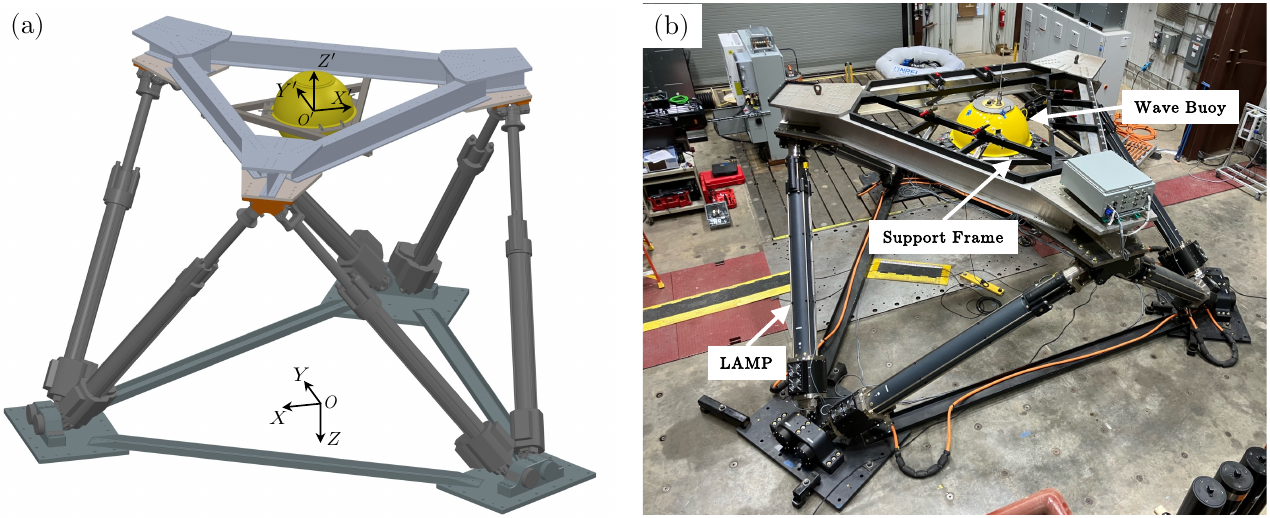}
    \caption{Experimental setup: (a) Diagram of experimental system and coordinate definition. Two different coordinates are used in the present study with $O-XYZ$ as LAMP motion coordinate and $O^\prime-X^\prime Y^\prime Z^\prime$ as buoy motion coordinate. (b) Overview of experimental setup. Buoy is fixed in LAMP system through a hex support frame.}
  \label{exppic}
\end{figure}

\cref{exppic} shows the experimental apparatus. A DWR-MkIII buoy (series number 73005) was mounted at the center of the platform and the platform coordinate origin via a hexagonal support frame. The measurable heave period range is \SI{1.6}{s} to $\SI{30}{s}$. The buoy's hull diameter is approximately $\SI{0.9}{m}$ and it has a mass of $\SI{216}{kg}$. The buoy provided wave elevation measurements at $\SI{1.28}{Hz}$ \citep{Datawell2025WaveriderManual}.

The six DOFs are defined in \cref{dofdefinition} based on the LAMP and buoy coordinates shown in \cref{exppic}. There are two perpendicular directional descriptions in the Datawell system: positive heave motion is defined as $-Z^\prime$ and positive vertical motion is defined as $Z^\prime$. The elevation $\eta$ is in the present study is defined as the same as vertical DOF.

\begin{table}[!h]
\centering
\caption{Definition of six degrees of freedom based on LAMP and buoy coordinate shown in \cref{exppic}. The positive rotation is defined by the right-hand rule.}
\begin{tabular}{@{\quad}ccc@{\quad}}
\toprule
DOF& LAMP Coordinate & Buoy Coordinate      \\ \midrule
Surge & $-X$            & $X^\prime$           \\
Sway  & $Y$             & $Y^\prime$           \\
Heave & $Z$             & $-Z^\prime$          \\
Vertical/Elevation & $-Z$ & $Z^\prime$  \\
Roll  & $-\theta_{X}$   & $\theta_{X^\prime}$  \\
Pitch & $\theta_{Y}$    & $\theta_{Y^\prime}$  \\
Yaw   & $\theta_{Z}$    & $-\theta_{Z^\prime}$ \\ \bottomrule
\end{tabular}
\label{dofdefinition}
\end{table}

Three signal pipelines were used: (1) the buoy radio chain, (2) the buoy internal chain, and (3) the LAMP optical motion chain. In the buoy radio chain, the Datawell Wave5 system communicated with the buoy via the HF link through Waverider Receiver RX-C4 and recorded the processed buoy output at \SI{1.28}{Hz}. In the buoy internal chain, the Buoy Tester software interfaced with the buoy over an RS-232 port to record the raw accelerometer signal at \SI{1.28}{Hz}. Within the buoy, acceleration is double-integrated and further processed to yield the elevation signal which is recorded through Wave5. Finally, in the LAMP optical motion chain, a Qualisys optical motion tracking system recorded the prescribed motion at \SI{100}{Hz}.

\subsection{Experiment matrix determination}
The prescribed motion is defined as
\begin{equation}
  \eta_m(t) = A_m \sin\!\left(\frac{2\pi t}{T_m}\right),
\end{equation}
where $A_m$ is the wave amplitude and $T_m$ is the wave period. The commanded motion $\eta_m$ is recorded with a Qualisys optical motion tracking system. $\eta_b$ denotes the corresponding buoy output signal from the Wave5 software. The prescribed motion is applied in the vertical direction.

The wave height measurement error is defined as
\begin{equation}\label{wheeq}
   \sigma = \frac{\eta_{m,\mathrm{rms}}-\eta_{b,\mathrm{rms}}}{\eta_{m,\mathrm{rms}}},
\end{equation}
where $\eta_{(-),rms}$ represents the root-mean-square value of the corresponding measurement signal.

The test matrix is constructed from four groups of test cases. The first is a uniformly sampled set of period–amplitude pairs with $T_m \in \{5,10,15,20,25,30\}\,\si{\second}$ and $A_m \in \{0.125,0.25,0.5,0.75\}\,\si{\meter}$. Second, Bayesian optimization is used to automatically select five additional test cases via the Adaptive Computing framework \citep{griffin2025adaptive}. The third part comprises eight additional cases with a fixed amplitude $A_m=\SI{0.25}{m}$ and varying periods $T_m \in \{1.6,\,2,\,4,\,12,\,22,\,24,\,25.5:0.5:29.5\}\,\si{\second}$, designed to cover the buoy’s measurable period range. In addition, three DWR-MkIII buoy datasets from real sea conditions, each spanning \SI{30}{min}, were reprocessed and replayed as prescribed motion inputs to evaluate the wave energy flux estimation error. The datasets were collected at three wave energy test sites: the U.S. Navy Wave Energy Test Site (WETS), PacWave, and Jennette’s Pier. All datasets are hosted by CDIP. In total, 37 harmonic prescribed motion tests were conducted, and three field tests were reprocessed, yielding 40 cases in the present study.

In the experiment, the longest case ran for more than \SI{30}{min}. A densely and uniformly sampled test matrix would therefore have resulted in a prohibitively long total measurement time. To increase test efficiency, i.e., to reduce the total test time, we employ a computer-assisted selection of test cases based on Bayesian optimization. The quantity of interest is the normalized error $\sigma$, which is modeled as a function of the two test matrix parameters $(T_m,A_m)$ using a Gaussian process:
\begin{equation}
    \sigma\!\left(T_m, A_m\right) \sim \mathcal{GP}\!\left(
        m\!\left(T_m, A_m\right),\;
        k\!\left(\left(T_m, A_m\right), \left(T_m^\prime, A_m^\prime\right)\right)
    \right),
\end{equation}
where $m(\cdot)$ and $k(\cdot,\cdot)$ are the mean and kernel (covariance) functions, respectively.
Here, $\left(T_m^\prime, A_m^\prime\right)$ denotes an independent input from $\left(T_m, A_m\right)$.

Given $n$ observations, the inputs $\mathbf{X}$ and outputs $\mathbf{Y}$ are defined as
\begin{equation}
\mathbf{X} \;=\; \left[\,
    \left(T_m^{(1)},A_m^{(1)}\right),\,\dots,\,\left(T_m^{(n)},A_m^{(n)}\right)
\,\right]^{\top},\qquad
\mathbf{Y} \;=\; \left[\,
    \sigma^{(1)},\,\dots,\,\sigma^{(n)}
\,\right]^{\top}.
\end{equation}
Let the covariance matrix $\mathbf{K}\in\mathbb{R}^{n\times n}$ be
\begin{equation}
K_{ij} \;=\; k\!\left(\left(T_m^{(i)},A_m^{(i)}\right),\,\left(T_m^{(j)},A_m^{(j)}\right)\right),
\end{equation}
and for a new condition $\left(T_m^*,A_m^*\right)$ we define
\begin{equation}
\mathbf{k}_* \;=\; \left[\,
    k\!\left(\left(T_m^*,A_m^*\right),\left(T_m^{(1)},A_m^{(1)}\right)\right),\,\dots,\,
    k\!\left(\left(T_m^*,A_m^*\right),\left(T_m^{(n)},A_m^{(n)}\right)\right)
\,\right]^{\top}.
\end{equation}
Assuming measurement noise with variance $\sigma_n^2$, the $\mathcal{GP}$ posterior at $\left(T_m^*,A_m^*\right)$ is Gaussian with mean and variance
\begin{equation}
\mu_*\!\left(T_m^*,A_m^*\right)
\;=\; m\!\left(T_m^*,A_m^*\right)
\;+\; \mathbf{k}_*^{\top}
\left(\mathbf{K}+\sigma_n^2 \mathbf{I}\right)^{-1}
\left(\mathbf{Y}-m\!\left(\mathbf{X}\right)\right),
\end{equation}
\begin{equation}
s_*^2\!\left(T_m^*,A_m^*\right)
\;=\; k\!\left(\left(T_m^*,A_m^*\right),\left(T_m^*,A_m^*\right)\right)
\;-\; \mathbf{k}_*^{\top}
\left(\mathbf{K}+\sigma_n^2 \mathbf{I}\right)^{-1}
\mathbf{k}_*.
\end{equation}
The hyperparameters of $m$ and $k$ (and $\sigma_n^2$) are obtained by maximizing the marginal likelihood. We adopt the maximum variance as the target for the next experiment by
\begin{equation}\label{gpvirance}
(T_m, A_m)_{next} = \underset{(T_m, A_m)}{\argmax}\; s_*(T_m, A_m),
\end{equation}
which represents the most uncertain region of the current $\mathcal{GP}$ model and efficiently refines the response surface $\sigma(T_m,A_m)$ with minimal trials. Additionally, the constraint $T_m>\SI{20}{s}$ was also imposed to target the high-period error region in the present study. The detailed iterations in the present study are shown in \ref{appendixgp}.

\section{Wave energy flux estimation method}\label{enemethodsec}
For the marine energy community, wave energy flux is the primary quantity of interest, not significant wave height $H_s$, which is often used in the physical oceanographic field. Several methods can be used to estimate wave energy from the time-domain wave elevation signal. We consider four methods in this section and note that only \cref{freene} in \cref{fremethodsec}, is included in the IEC~62600 technical specifications.

\subsection{Frequency-domain method}\label{fremethodsec}
\begin{figure}[ht!]
    \centering
    \includegraphics[width=.5\textwidth]{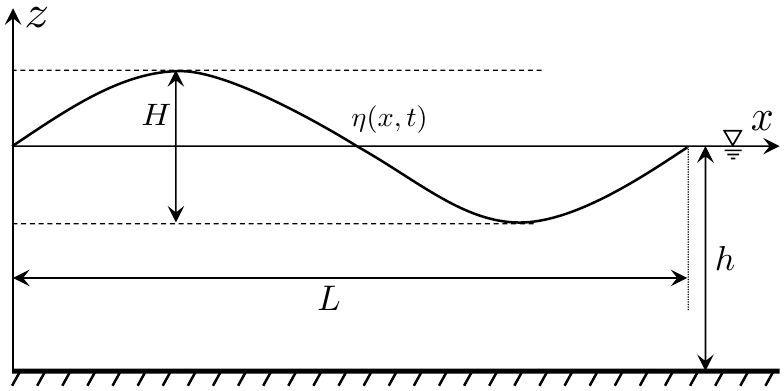}
    \caption{Definition of wave parameters. $h$ is the water depth, $L$ is the wave length, $H$ is the wave height, and $\eta(x,t)$ is the wave elevation.}
  \label{wavestrupic}
\end{figure}

In the IEC~62600-100 technical specification, the omnidirectional wave energy flux $J_f$ is defined as:
\begin{equation}\label{freene}
J_f=\rho g\sum_iS_ic_{gi}\Delta f_i ,
\end{equation}
where $J_f$ represents the energy flux (\si{W/m}) estimated from a wave spectrum $S_i$, the $i$th component of the wave spectrum density ($\si{m^2/Hz}$), $c_{gi}$ is the corresponding group speed, and $\Delta f_i$ is the $i$th frequency width. $S$ is the power spectral density (PSD) calculated from the wave elevation. The IEC~62600 technical specification does not specify a PSD estimation method, so the Welch PSD estimation method is applied as shown in Algorithm~\ref{alg:welch_psd}. This method differs from that stated in the Datawell manual \citep{Datawell2025WaveriderManual}, which involves additional smoothing procedures in the processing. Any systematic differences between these two methods will be explored in future work. The PSD estimation method described in Algorithm~\ref{alg:welch_psd} with $50\%$ overlap and $N_{FFT} = N_w = 256$ is applied. $c_{gi}$ is the group velocity defined as
\begin{equation}\label{groupveleq}
    c_{g_i} = \frac{1}{2} c_{p_i} \left[ 1 + \frac{2k_i h}{\sinh(2k_i h)} \right],
\end{equation}
where the phase velocity $c_{p_i}$ is
\begin{equation}
    c_{p_i} = \sqrt{\frac{g}{k_i} \tanh(k_i h)},
\end{equation}
where $g = \SI{9.8}{m/s^2}$ is the gravitational acceleration, $k_i$ is the wavenumber, and $h$ is the water depth, which is chosen as $\SI{70}{m}$ in the present study because it is the water depth at the PacWave wave energy testing site, as shown in \cref{wavestrupic}. The wavenumber $k_i$ is obtained by solving the linear dispersion relation by the Newton-Raphson method:
\begin{equation}
    \omega_i^2 = g k_i \tanh(k_i h),
\end{equation}
where $\omega_i = 2\pi f_i$ is the angular frequency of the corresponding bin. Then, the energy flux can be estimated through the frequency-domain method based on \cref{freene}.

\begin{algorithm}[ht!]
\caption{Welch method for power spectral density estimation}
\label{alg:welch_psd}
\KwIn{Wave elevation $x[1{:}L_x]$, sampling rate $f_s$, window length $L_w$, overlap $N_{overlap}$, oven FFT length $N_{FFT}$, window function $w[1{:}L_w]$ (Hanning window).}
\KwOut{Frequencies $f[1{:}M]$ and $\mathrm{PSD}[1{:}M]$ in [$\mathrm{m}^2/\mathrm{Hz}$].}

$N_{step} \leftarrow L_w - N_{overlap}$\;
$w[k] \leftarrow 0.5\Big(1 - \cos\big(\frac{2\pi (k-1)}{L_w-1}\big)\Big),\ k = 1{:}L_w$\tcp*[r]{Hanning window function}

$w[k] \leftarrow w[k]\Big/\Big(\frac{1}{L_w}\sum_{k=1}^{L_w} w[k]\Big)$\; 
$\mathbf{P}[1{:}N_{FFT}]\leftarrow 0$;\quad $N_{seg}\leftarrow 0$\;

$N_{seg} \leftarrow \left\lfloor \dfrac{L_x - N_{overlap}}{N_{step}} \right\rfloor$\; 

\For{$s \leftarrow 1$ \textbf{\textup{to}} $N_{seg}$}{
  $i_{\mathrm{start}} \leftarrow (s{-}1)N_{step} + 1$\;
  $x_s[j]\leftarrow x[i_{\mathrm{start}}{+}j{-}1],\ j=1{:}L_w$\;
  $x_s[j]\leftarrow x_s[j]-\dfrac{1}{L_w}\sum_{j=1}^{L_w}x_s[j]$\;
  $x_s[j]\leftarrow x_s[j]\cdot w[j]$\;
  $X_s \leftarrow FFT(x_s, N_{FFT})$\tcp*[r]{Fast Fourier Transform
}
  \For{$m \leftarrow 1$ \textbf{\textup{to}} $N_{FFT}$}{
    $\mathbf{P}[m]\leftarrow \mathbf{P}[m]+|X_s[m]|^2$\;
  }
}
$\mathbf{P}[m]\leftarrow \mathbf{P}[m]/N_{seg}$\;
$M \leftarrow \lfloor N_{FFT}/2 \rfloor$\;

$\mathbf{P}_{1\mathrm{s}}[1]\leftarrow \mathbf{P}[1]$; 

\For{$m \leftarrow 2$ \textbf{\textup{to}} $M$}{ $\mathbf{P}_{1\mathrm{s}}[m]\leftarrow 2\,\mathbf{P}[m]$ \tcp*[r]{one-sided spectrum}}

$\displaystyle \mathrm{PSD}[m]\leftarrow
\dfrac{\mathbf{P}_{1\mathrm{s}}[m]}{f_s\sum_{k=1}^{L_w}w[k]^2}$;\quad
$f[m]\leftarrow \dfrac{(m{-}1)}{N_{FFT}}f_s,\ m=1{:}M$\;
\end{algorithm}

\subsection{Time-domain method based on small-amplitude wave theory} 
The velocity potential function in small-amplitude wave theory (also Airy wave theory or linear wave theory) is:
\begin{equation}
    \Phi=\frac{gH}{2\omega}\frac{\cosh k(z+h)}{\cosh kh}\sin(kx-\omega t),
\end{equation}
and the wave energy over one wavelength in one period contains kinetic energy $E_k$ and potential energy $E_p$:
\begin{equation}
    \begin{aligned}
        E_k&=\int_0^L\int_{-h}^\eta\frac{1}{2}\rho(u_x^2+u_z^2)\mathrm{d}xdz = \dfrac{1}{16}\rho g H^2 L,\\
        E_{p}&=\int_0^L\frac{1}{2}\rho g(h+\eta)^2dx = \dfrac{1}{16}\rho g H^2 L.
    \end{aligned}
\end{equation}
The total energy per unit wave length is then $E = \dfrac{1}{8}\rho g H^2$ with units of $[\mathrm{J}/\mathrm{m^2}]$. The mean energy transfer within one period is
\begin{equation}
    \bar{E} = \dfrac{1}{T}\int_0^T\int_{-h}^\eta p u_xdzdt = Ec_g,
\end{equation}
where $c_g$ is the group velocity shown in \cref{groupveleq}. Then, the energy flux based on small-amplitude wave theory $J_s$ can be estimated as 
\begin{equation}\label{smallwaveeq}
   J_s = \dfrac{\sum_{i=1}^N \bar{E}_i\cdot T_i }{\sum_{i=1}^N T_i}. 
\end{equation}

\subsection{Time-domain method based on Hilbert transform}
The envelope of the wave elevation $A(t)$ can be expressed as \citep{Hwang2002EnergyFluxHilbertWavelet}
\begin{equation}
\begin{aligned}
L(t)&=\eta(t)+i\mathcal{H}[\eta(t)]=A(t)e^{i\phi(t)},\\
A(t) &= \sqrt{\eta(t)^2+(\mathcal{H}[\eta(t)])^2},
\end{aligned}
\end{equation}
where $\mathcal{H}[-]$ is the Hilbert transform and $L(t)$ is the analytic signal. For wave elevation signals with multi-frequency components, empirical mode decomposition can be used to extract intrinsic mode functions. Applying the Hilbert transform to the intrinsic mode functions yields the Hilbert–Huang transform. However, since monochromatic wave motion is used in the present study, only the Hilbert transform is employed. The instantaneous wave energy can be expressed as 
\begin{equation}
    E(t) = \dfrac{1}{2}\rho g A^2(t),
\end{equation}
and the wave energy flux based on the Hilbert transform can be expressed as
\begin{equation}
    J_h =\dfrac{1}{t_{max}}\int_0^{t_{max}}\dfrac{1}{2}\rho g A^2(t)c_g dt,
\end{equation}
where $c_g$ is the group velocity given in \cref{groupveleq}.

\subsection{Time-domain method based on wavelet transform}
Wavelet analysis \citep{lueck2000wavelet} can be expressed in the frequency domain based on Parseval’s theorem as
\begin{equation}\label{waveletantoms}
\begin{aligned}
    \mathcal{WT}(\tau,s)&=\frac{1}{\sqrt{s}}\int_{-\infty}^{\infty}\eta(t)\psi^*_{\tau,s}\left(t\right)dt = \frac{1}{\sqrt{s}}\int_{-\infty}^{\infty}\eta(t)\psi^*\left(\frac{t-\tau}{s}\right)dt\\
    &=\dfrac{1}{2\pi}\int_{-\infty}^{\infty}H(\omega)\Psi^*_{\tau,s}(\omega)d\omega,
\end{aligned}
\end{equation}
where $(-)^*$ denotes a complex conjugate, and 
\begin{equation}
    \psi_{\tau,s}(t) = \frac{1}{\sqrt{s}}\psi\left(\frac{t-\tau}{s}\right)
\end{equation}
are the wavelet atoms constructed by a time translation $\tau$, scale dilation $s$, and the mother wavelet function $\psi(t)$. The Fourier transform of wavelet atoms is: 
\begin{equation}
    \Psi_{\tau,s}(\omega) = \int_{-\infty}^{\infty}\psi_{\tau,s}(t)e^{-i\omega t} dt = \sqrt{s}e^{-i\omega \tau}\Psi(s\omega).
\end{equation}

Considering the convolution theorem, the Fourier transform of the wavelet transform $\mathcal{WT}(\tau,s)$ can be expressed as \citep{huang2004wave}:
\begin{equation}
    \mathcal{F}\left[\mathcal{WT}(\tau,s)\right]=\mathcal{F}\left[ \frac{1}{\sqrt{s}}\int_{-\infty}^{\infty}\eta(t)\psi^*\left(\frac{t-\tau}{s}\right)dt \right]=H(\omega)\sqrt{s}\Psi^*(s\omega),
\end{equation}
where $\mathcal{F}[-]$ is the Fourier transform. The wavelet transform is then computed as
\begin{equation}
    \mathcal{WT}(\tau,s) = \mathcal{F}^{-1}\left[H(\omega)\sqrt{s}\Psi^*(s\omega)\right],
\end{equation}
and the wavelet power spectrum is defined as:
\begin{equation}
    P_{\mathcal{WT}}(t,s) = \lvert\mathcal{WT}(\tau,s)\rvert^2.
\end{equation}

The Morlet wavelet is chosen as the mother wavelet:
\begin{equation}
    \psi(t)=\exp(\frac{-t^2}{2})\mathrm{exp}(ict),
\end{equation}
$c$ is set as $6.00$ \citep{huang2004wave}, and the admissibility coefficient is
\begin{equation}
    C_{\psi} = \int_0^\infty\frac{|\Psi(\omega)|^2}{\omega}d\omega = 1.883.
\end{equation}
Then, the wave energy spectrum can be expressed as 
\begin{equation}
    S(t,f)=\frac{2}{C_\psi s f}|\mathcal{WT}(\tau,s)|^2,
\end{equation}
the boundary energy spectrum is 
\begin{equation}
    E(t)=\int S(t,f)df,
\end{equation}
and the energy flux based on the wavelet transform can be expressed as
\begin{equation}\label{waveleteq}
    J_w= \dfrac{\sum E(t)\rho g c_g(f)}{t_{max}}.
\end{equation}

MATLAB Wavelet Toolbox releases after 2017 provide only $L1$ normalized wavelets and do not return normalized coefficients; therefore, they could not produce the energy flux estimation directly in the present study.

\section{Results and discussion}
In this section, we present the direct measurement errors, wave height errors, and the corresponding errors in wave energy flux estimation. The wave height errors are obtained from the experiment described in \cref{expsec}, and the energy flux is further estimated using the methods described in \cref{enemethodsec}.

\subsection{Wave height measurement error}\label{heighterror}
\begin{figure}[ht!]
    \centering
    \includegraphics[width=.7\textwidth]{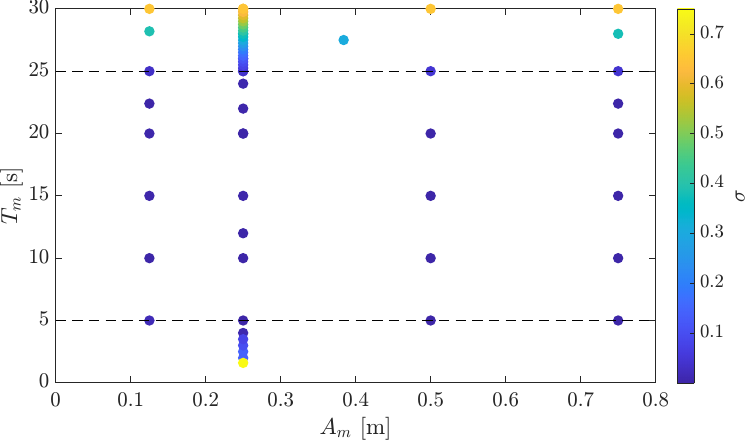}
    \caption{Distribution of the wave height measurement error $\sigma$ over the amplitudes $A_m$ and periods $T_m$. Each dot corresponds to one test condition, and the color indicates the value of $\sigma$ shown in the colorbar.}
  \label{heightall}
\end{figure}

\begin{figure}[ht!]
    \centering
    \includegraphics[width=.99\textwidth]{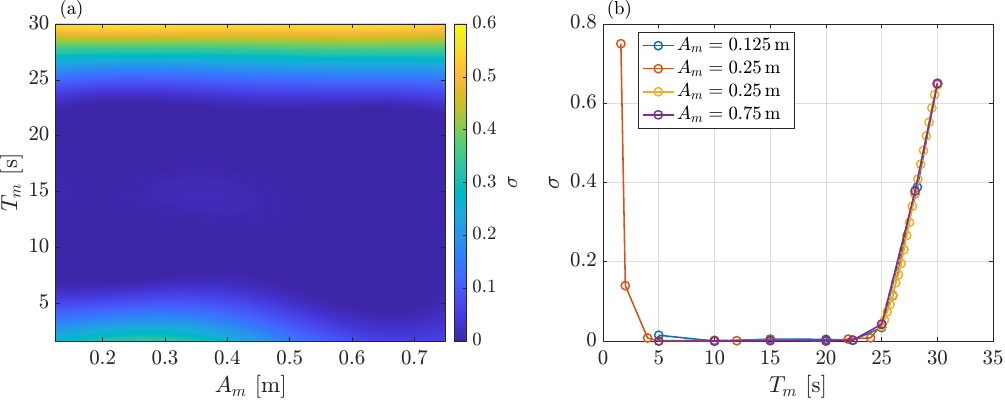}
    \caption{Contour and scatter results of the wave height measurement error $\sigma$: (a) Contour distribution of wave height measurement $\sigma$ over $T_m$ and $A_m$; (b) variation of $\sigma$ with $T_m$ for several amplitudes $A_m$. The error increases in both short- and long-period ranges, consistent with the discrete results shown in \cref{heightall}.}
  \label{diffallpic}
\end{figure}

\cref{heightall} and \cref{diffallpic} show the results of testing over $T_m$ and $A_m$ with the difference between the platform heave motion and the buoy output calculated based on \cref{wheeq}. \cref{heightall} plots the error for each case as separate points where the color of the points indicates the magnitude of error. In \cref{diffallpic}(a), a colormap is created by interpolating the grid of wave height errors where the color indicates the value of $\sigma$. Subfigure(b) plots the error for fixed amplitudes $A_m$. The acceleration limits of the motion platform precluded the combinations of high $A_m$ and low $T_m$, which resulted in inconsistencies in the error contours at short periods across different $A_m$ in \cref{diffallpic}(a). The error is below $1\%$ for most test cases. The error rapidly increases at short (low) period waves ($T_m < \SI{5}{s}$) and long (high) period waves ($T_m > \SI{25}{s}$), which are the measurement limits of the wave buoy. The maximum error in the long-period range and short-period range is greater than $60\%$. There is a consistent relationship between the error $\sigma$ and $T_m$ across different $A_m$. In the long-period region, the error is essentially independent of $A_m$. In the short-period region, the error is evaluated at the fixed amplitude $A_m=\SI{0.25}{m}$, as shown in \cref{diffallpic}(b).

\begin{figure}[ht!]
    \centering
    \includegraphics[width=.75\textwidth]{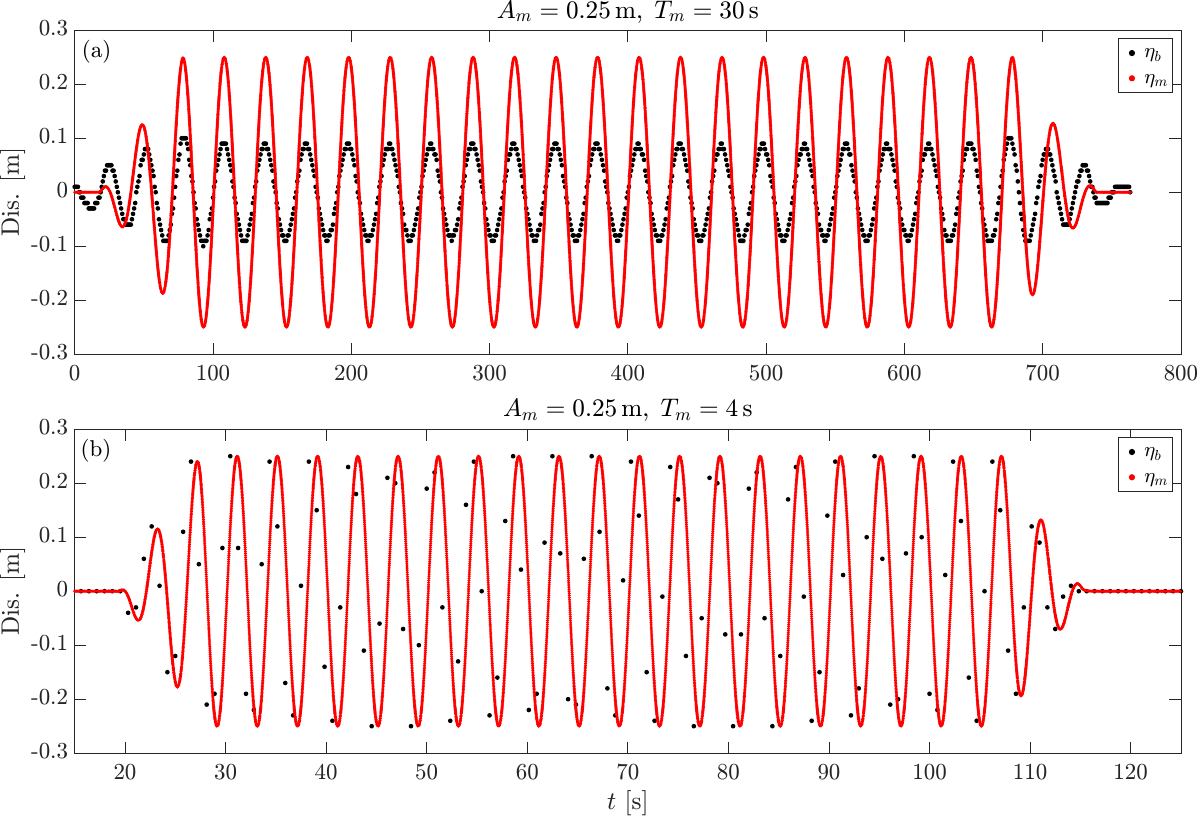}
    \caption{Comparison of prescribed and measured elevation time series for two representative cases at fixed $A_m=\SI{0.25}{m}$ and different $T_m$: (a) Long-period case with $T_m = \SI{30}{s}$; (b) short-period case with $T_m=\SI{4}{s}$. The red line denotes the prescribed motion $\eta_m$ with sampling frequency of $\SI{100}{Hz}$, and the black dots denote the buoy output signal $\eta_b$ with sampling frequency of $\SI{1.28}{Hz}$.}
  \label{elevpic}
\end{figure}

The time series of the buoy output and prescribed motion are shown in \cref{elevpic}, where the prescribed platform motion and the measured wave elevation signals are overlayed for two representative cases at the fixed amplitude $A_m=\SI{0.25}{m}$ and periods $T_m = \SI{30}{s}$ and $\SI{4}{s}$, respectively. For the long-period case, the buoy output elevation $\eta_b$ is consistently lower in amplitude than the prescribed motion $\eta_m$ by an approximately constant factor, as \cref{elevpic}(a) shows. There are two possible reasons for this: one is underestimation in the raw acceleration measurement, and the other is additional postprocessing, such as analog or digital filters, applied during the double-integration procedure. 

\begin{figure}[ht!]
    \centering
    \includegraphics[width=.75\textwidth]{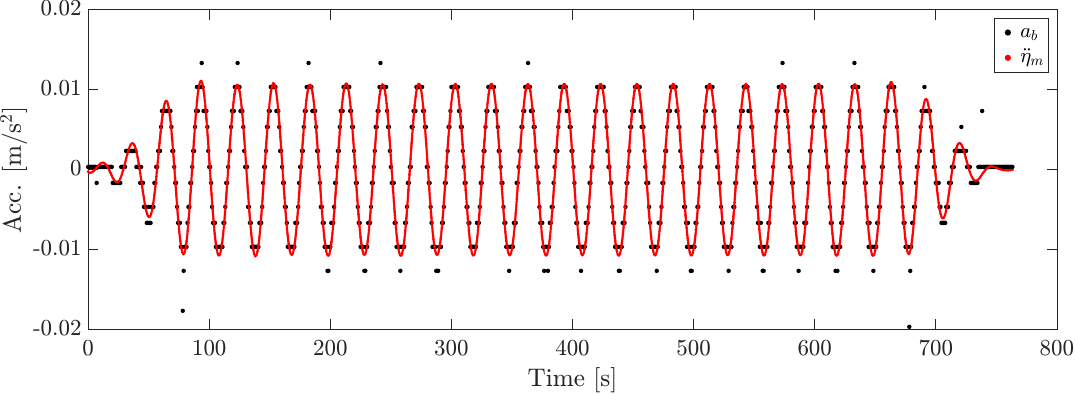}
    \caption{Comparison of prescribed and measured acceleration time series for the case of $A_m=\SI{0.25}{m}$, $T_m=\SI{30}{s}$. Raw acceleration from the Buoy Tester software is shown as black dots, and red dots represent the Qualisys acceleration.}
  \label{accpic}
\end{figure}

\cref{accpic} compares the acceleration signals of the prescribed and measured motions. The red markers denote the acceleration obtained by twice differentiating in time the displacement measured by the Qualisys motion-tracking system. The black dots represent the raw acceleration signal from the buoy. The results show that the buoy acceleration tracks the prescribed motion but slightly overshoots the prescribed peaks and troughs in acceleration, indicating the consistency of the acceleration measurement. Consequently, any deviation in the elevation signal arises from additional internal postprocessing performed in the double-integration procedure.

For the short-wave-period case, the buoy output elevation matches the prescribed motion amplitude, as shown in \cref{elevpic}(b). However, the signal $\eta_b$ exhibits apparent amplitude beating. The Nyquist frequency of the buoy is $\SI{0.64}{Hz}$, and the prescribed motion in this case is at $\SI{0.25}{Hz}$. The beating phenomenon is observed despite the motion frequency being below the Nyquist limit, which is known as a sub-Nyquist artifact \citep{amidror2015sub}. Consequently, the wave height measurement error in the short-period region arises from a sub-Nyquist artifact associated with the relatively low sampling frequency of \SI{1.28}{Hz}.

In conclusion, there are two error regions, one at either limit of the specified wave buoy measurement range  (\SIrange{1.6}{30}{s}). For the long-period region, $T_m > \SI{25}{s}$, the maximum wave height error reaches approximately $65\%$, attributable to additional internal postprocessing applied in the double-integration procedure. And for the short-period region, $T_m < \SI{5}{s}$, a sub-Nyquist artifact arises, producing measurement error. The error depends primarily on period rather than amplitude and remains less than $1\%$ across the remaining period range for monochromatic waves.

\subsection{Wave energy flux estimation error}
Wave energy error is analyzed using the wave energy flux estimation methods that were introduced in \cref{enemethodsec}.

\subsubsection{Short-period wave energy flux estimation}

For short-period waves, $T_m < \SI{5}{s}$, the sub-Nyquist artifact occurs and influence the wave height measurement accuracy. To identify the influence of wave height measurement errors on the estimated wave energy flux, we further investigate the theoretical energy flux $J$ computed using the four methods described in \cref{enemethodsec}. $\Xi$ is defined as the energy flux error: 
\begin{equation}\label{enediffeq}
    \Xi = \dfrac{J_m-J_b}{J_m},
\end{equation} 
where $J_m$ is the wave energy flux computed from the prescribed platform motion sampled at $\SI{100}{Hz}$, and $J_b$ is the wave energy flux computed from the buoy output sampled at $\SI{1.28}{Hz}$.

Short-period waves are prone to wave breaking under certain sea states because their larger steepness makes them dynamically unstable. Two wave breaking criteria are applied to identify the possible wave period and height combinations that are unrealistic in the real ocean:
\begin{enumerate}
    \item Shallow water effect:
    \begin{enumerate}
        \item \emph{Deep water} $(h/L > 0.5)$: flag if
    \[
      \frac{H}{L} > 0.142.
    \]
        \item \emph{Shallow water} $(h/L < 0.05)$: flag if
    \[
      \frac{H}{d} > 0.78.
    \]
    \end{enumerate}
    \item Wave steepness effect \citep{perlin2013breaking}: flag if
    \[
      \dfrac{kH}{2}>0.443.
    \]    
\end{enumerate}

\begin{figure}[ht!]
    \centering
    \includegraphics[width=.5\textwidth]{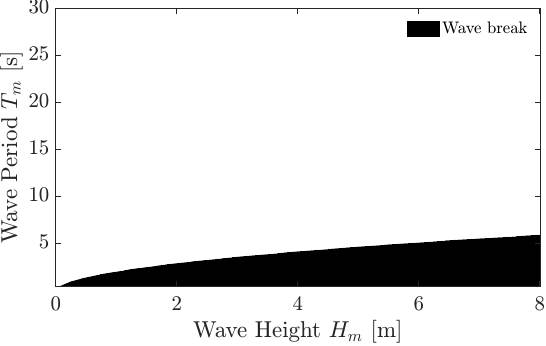}
    \caption{Wave breaking region in the wave height and period plane. Black shaded area indicates combinations of wave height and period where breaking waves are likely.}
  \label{wavebreakregion}
\end{figure}

\begin{figure}[ht!]
    \centering
    \includegraphics[width=.99\textwidth]{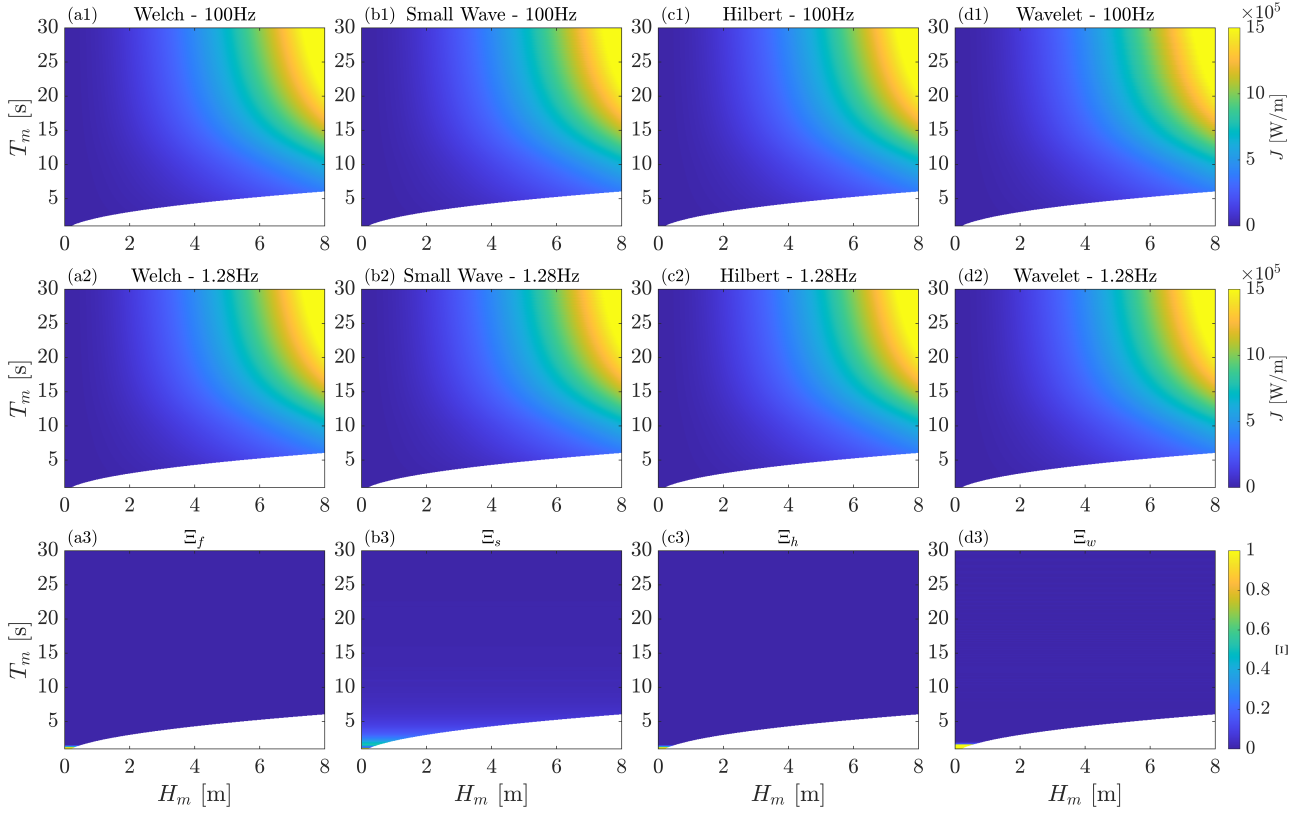}
    \caption{Comparison of wave energy flux and corresponding relative errors across different methods and sampling frequencies. Subfigure(a1) to (d1) show the energy flux calculated using each of the four estimation methods stated in \cref{enemethodsec} and using the prescribed elevation data as measured by the Qualisys motion capture system. Subfigures(a2) to (d2) show the corresponding values calculated using the buoy's output elevation data. Subfigures(a3) to (d3) present the energy flux estimation error $\Xi$ based on different methods by \cref{enediffeq}. The white regions indicate wave breaking occurrence.}
  \label{lowTerror}
\end{figure}

\cref{wavebreakregion} shows the wave breaking region across different wave height and period combinations; the black area indicates conditions under which wave breaking is likely in the ocean. \cref{lowTerror} presents the theoretical wave energy flux estimation result based on different methods across combinations of wave periods and heights, with the wave breaking area shown in white. \cref{lowTerror}(a1) to (d1) present the wave energy flux colormaps computed from the prescribed wave elevation signal measured using the Qualisys system at $\SI{100}{Hz}$ and calculated using the frequency-domain method, small-amplitude wave theory, the Hilbert transform, and the wavelet transform, respectively. Subfigure(a2) to (d2) present the corresponding wave energy flux calculated using the buoy's measured wave elevation signal at $\SI{1.28}{Hz}$. Subfigures(a3) to (d3) present the wave energy flux errors as defined in \cref{enediffeq}, for four methods: $\Xi_f$ (frequency-domain method), $\Xi_s$ (small amplitude wave theory), $\Xi_h$ (Hilbert transform), $\Xi_w$ (wavelet transform). 

As shown in \cref{lowTerror}(a1) to (d1), the wave energy flux $J$ increases markedly with increasing wave height $H_m$ and period $T_m$. For all methods, the high-energy region concentrates toward the upper right area of each panel (large $H_m$, long $T_m$) while the low-energy region lies near the lower-left (small $H_m$, short $T_m$). Energy flux $J$ contours are smooth and mutually consistent across methods, clearly delineating a monotonic growth of energy with $H_m$ and $T_m$. The wave energy flux estimates computed at the low sampling rate of $\SI{1.28}{Hz}$ exhibit results similar to those obtained at the high sampling frequency, as \cref{lowTerror}(a2) to (d2) show. \cref{lowTerror}(a3) to (d3) present the wave energy flux estimation error for different methods based on \cref{enediffeq}. For most combinations of period and amplitude with $T_m>\SI{5}{s}$, the error is below $2\%$, whereas larger errors occur in the short-period region.

\begin{figure}[ht!]
    \centering
    \includegraphics[width=.99\textwidth]{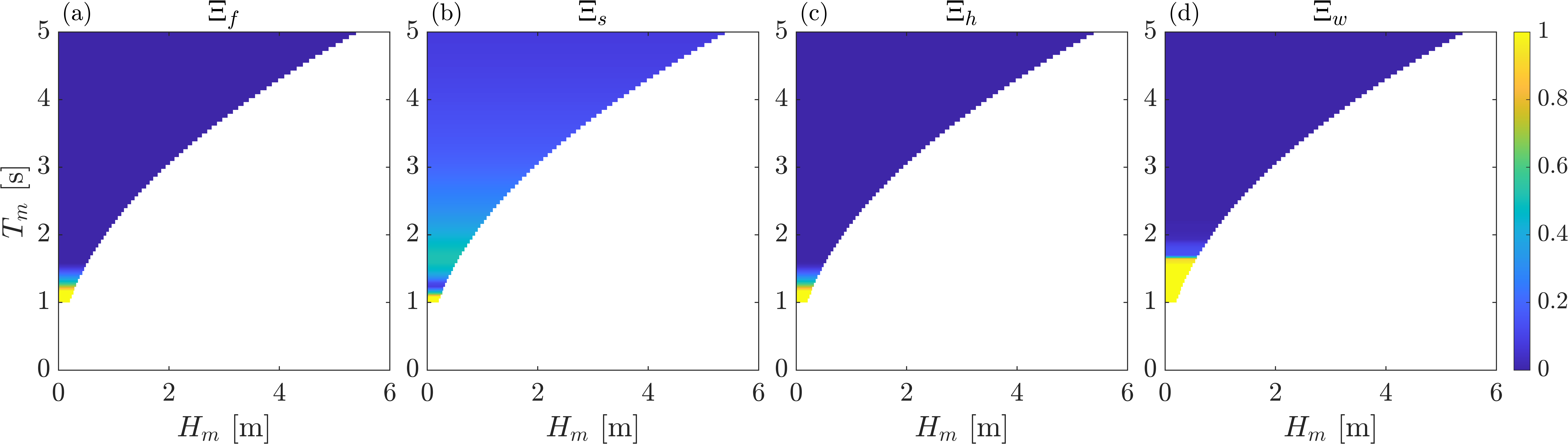}
    \caption{Energy flux estimation error $\Xi$ for cases with $T_m<\SI{5}{s}$, computed using different methods as defined in \cref{enediffeq}: (a) frequency-domain method, (b) small-amplitude wave theory, (c) Hilbert transform, (d) wavelet transform. The white regions indicate wave breaking occurrence.}
  \label{lowTerror0to5}
\end{figure}

\cref{lowTerror0to5} presents the wave energy flux error in the short-period range $T_m < \SI{5}{s}$ zoomed in on the region of largest error. In all subplots, the wave energy flux estimation error increases rapidly when $T_m<\SI{2}{s}$. The error peak is above $100\%$ near $T_m\approx\SI{1}{s}$. These results indicate that the low sampling frequency of \SI{1.28}{Hz} can introduce wave energy flux estimation errors exceeding $100\%$ for wave periods shorter than $\SI{2}{s}$, whereas the estimates become reliable for $T_m>\SI{5}{s}$. It should be noted that there is no established best practice for estimating wave energy flux in this range of wave periods and low sampling frequencies. This will be investigated in future studies. Here, we show that energy flux estimation errors are significant in the short-period range at this sampling rate and should be carefully considered in corresponding wave resource characterization.

\subsubsection{Long-period wave energy flux estimation}
\begin{figure}[ht!]
    \centering
    \includegraphics[width=.99\textwidth]{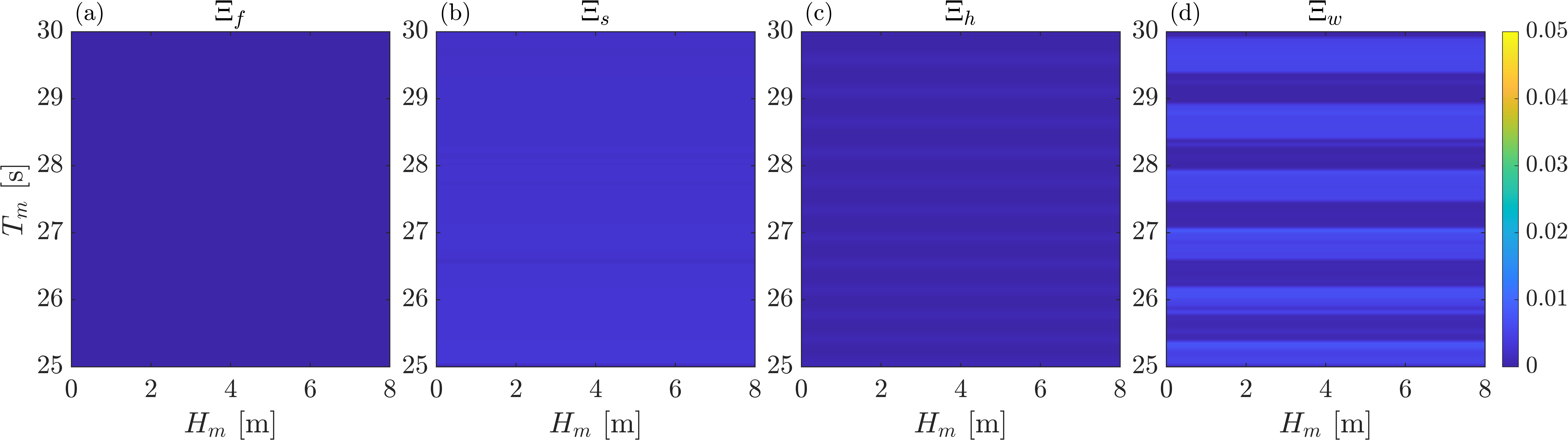}
    \caption{Energy flux estimation error $\Xi$ for cases with $T_m>\SI{25}{s}$, computed using different methods as defined in \cref{enediffeq}: (a) frequency-domain method, (b) small-amplitude wave theory, (c) Hilbert transform, (d) wavelet transform.}
  \label{highTerror25to30}
\end{figure}

\cref{highTerror25to30} shows the detailed distribution of error for the long-period wave range $\SI{25}{s} < T_m < \SI{30}{s}$. The results confirm that the wave energy flux estimation error calculated between the four methods and the wave buoy measurements is less than $2\%$. Thus, the low sampling frequency of $\SI{1.28}{Hz}$ does not impact the energy flux estimation for the period range $\SI{25}{s} < T_m < \SI{30}{s}$. Therefore, the main source of error within this wave period range is the wave elevation error introduced by additional internal postprocessing shown in \cref{elevpic}(a).

As stated in \cref{highTerror25to30}, for longer-period waves, $T_m > \SI{25}{s}$, there is no sub-Nyquist artifact, and the sampling frequency of $\SI{1.28}{Hz}$ does not influence the energy flux, as \cref{lowTerror} shows. 

Considering small-amplitude wave theory, the wave energy flux $J$ scales with the square of wave height
\begin{equation}
    J\propto H_m^{2} \propto \eta^2_{(-),rms}.
\end{equation}
Revisiting \cref{enediffeq}, the wave height measurement error $\sigma$ can also be expressed as $\eta_{b,rms} = (1-\sigma)\eta_{m,rms}$; then, the relative energy error $\Xi := (J_m-J_b)/J_m$ can be expressed as
\begin{equation}\label{eneerroreq}
    \Xi = \dfrac{J_m-J_b}{J_m} = \dfrac{J_m-(1-\sigma)^2J_m}{J_m} = 2\sigma - \sigma^2.
\end{equation}

\begin{figure}[ht!]
    \centering
    \includegraphics[width=.85\textwidth]{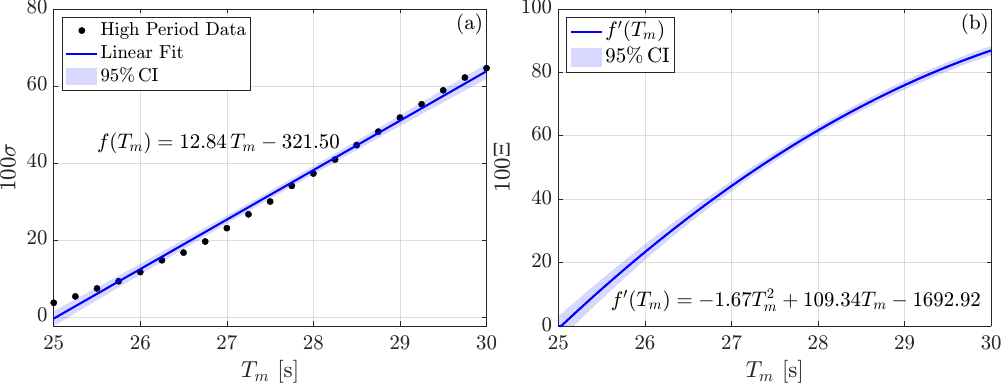}
    \caption{Regression analysis of the wave energy flux estimation error at long wave periods: (a) Linear fitting $f(T_m)$ of wave height measurement error $\sigma$ with respect to period $T_m$. The $100\sigma$ is analyzed for better illustration with the formula $100\sigma = 12.81 T_m - 321.50$; (b) energy estimation error $f(T_m)$ calculated by \cref{eneerroreq} as $f^\prime (T_m) = -1.67 T_m^2 + 109.34 T_m -1692.92$. The light blue area represents the 95\% confidence interval.}
  \label{higherror25T30}
\end{figure}

As shown in \cref{higherror25T30}, both the wave height measurement error $\sigma$ and the energy flux estimation error $\Xi$ increase with increasing wave period $T_m$ in the long-period region. In subfigure(a), the linear regression of $100\sigma$ yields $100\sigma = 12.81 T_m - 321.50$, suggesting a nearly proportional relationship between wave measurement uncertainty and wave period. Subfigure(b) shows the corresponding energy flux estimation error computed by \cref{eneerroreq}, which follows a polynomial relation $100\Xi = -1.67 T_m^2 + 109.34 T_m -1692.92$. The nonlinear growth of $\Xi$ with $T_m$ suggests that long-period waves contribute disproportionately to the overall energy estimation error. Together, these results demonstrate that long-period waves like swell, although energetically dominant, can cause energy estimation losses exceeding $80\%$ when measured with a DWR-MkIII buoy. \citet{van2018wave} reported buoy measurement errors in the long-period range, and the present study quantifies these errors for the first time. The impact of wave-height measurement errors on the energy flux estimates will be investigated in future work.

\subsection{Wave energy testing site cases}
In this section, we introduce the wave energy flux estimation error based on three testing site cases. The CDIP datasets are reprocessed to reconstruct the prescribed buoy motions, and the corresponding buoy outputs are subsequently analyzed. Three representative cases are as follows:
\begin{enumerate}
    \item WETS \citep{cross2020recent}: 30 min data from 19:00 to 19:30, Jan.~4,~2025, of 225pz buoy in CDIP system\footnote{https://cdip.ucsd.edu/m/products/?stn=225pz}.
    \item PacWave \citep{dunkle2020pacwave}: 30 min data from 21:00 to 21:30, June~28,~2024, of 277p1 buoy in CDIP system\footnote{https://cdip.ucsd.edu/m/products/?stn=277p1}. 
    \item Jennette’s Pier \citep{osti_1963664}: 30 min data from 19:00 to 19:30, Aug.~16,~2023, of 243p1 buoy in CDIP system\footnote{https://cdip.ucsd.edu/m/products/?stn=243p1}. 
\end{enumerate}

\begin{figure}[ht!]
    \centering
    \includegraphics[width=.99\textwidth]{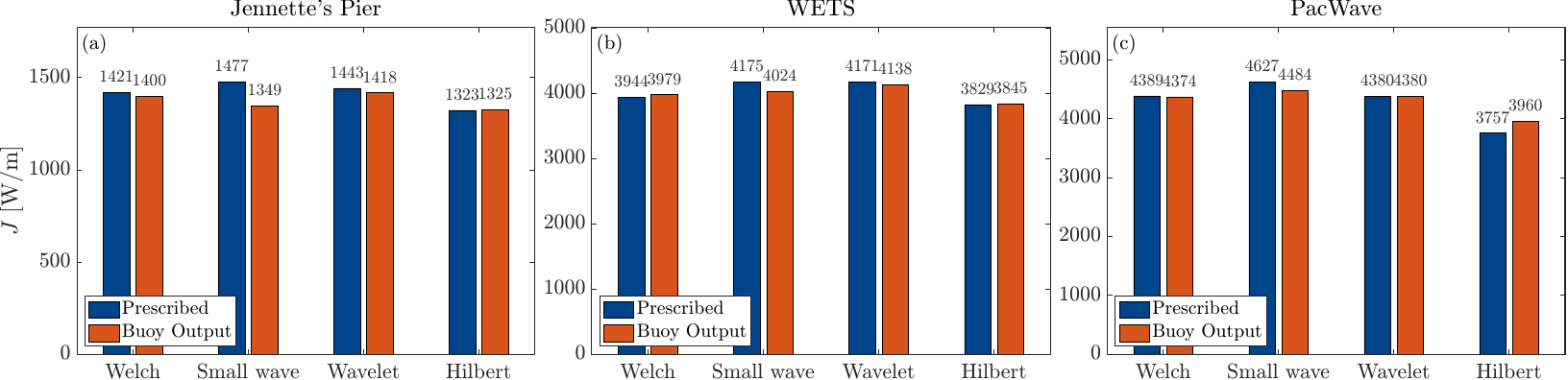}
    \caption{Comparison of the wave energy flux $J$ estimation at three wave energy testing sites based on different analysis methods: (a) Jennette’s Pier site, (b) WETS site, and (c) PacWave site. Each subfigure compares the prescribed reference wave energy flux and the buoy output flux obtained using four methods: frequency-domain method (Welch), small-amplitude wave theory (Small wave), Wavelet transform (Wavelet), and Hilbert transform (Hilbert). The prescribed motion is at $\SI{100}{Hz}$, and the buoy output motion is at $\SI{1.28}{Hz}$.}
  \label{irreguene}
\end{figure}

\cref{irreguene} shows the comparison results of wave energy flux estimation of three different wave energy testing sites. Four different methods are applied with prescribed motion at $\SI{100}{Hz}$ and buoy output motion at $\SI{1.28}{Hz}$. The wave energy flux calculated using the frequency-domain method based on \cref{freene} at $\SI{100}{Hz}$ is used as a baseline because it is specified in IEC Technical Specification 62600 \citep{IEC62600-100-2024}. Energy flux estimations obtained with this method at different sampling frequencies agree within $2\%$ in these three cases. Among the four methods, the Hilbert transform method underestimates wave energy flux by $6.90\%$, $2.92\%$, and $14.40\%$ for Jennette’s Pier, WETS, and PacWave, respectively. Since real ocean waves contain multiple-frequency components, a pure Hilbert transform may not be sufficient, and the Hilbert-Huang transform will be analyzed in a future study. Compared with the frequen—overpredicthods, small-amplitude wave theory and the wavelet transform based on \cref{smallwaveeq,waveleteq}, overpredict wave energy flux by a maximum $\SI{50}{W/m}\;(3.52\%)$, $\SI{231}{W/m}\;(5.86\%)$, and $\SI{238}{W/m}\;(5.42\%)$ across the three datasets. In this study, it is shown that the frequency-domain method may introduce approximately $6\%$ wave energy flux estimation error for real testing site resource characterization. Further uncertainty quantification analysis of energy flux estimation methods is needed.

\section{Conclusion and future work}\label{conclu}

Prescribed motion experiments were conducted on a large-amplitude motion platform facility at the National Laboratory of the Rockies to quantify the heave measurement performance of a Datawell Waverider MkIII buoy and the resulting impact on wave energy flux. This work extends existing dry buoy calibration from a single fixed motion setting to the full operational range of wave periods. Forced motions were validated with the Qualisys optical system while buoy elevation and raw acceleration were recorded. The elevation errors were propagated to wave energy flux using four different calculations: a frequency-domain method, an approach based on small-amplitude wave theory, a Hilbert transform approach, and a wavelet transform approach. Three CDIP field records (Jennette’s Pier, WETS, and PacWave) were recreated to test practical performance. The main findings are:

1) The buoy measurements are accurate for wave periods between $\SI{5}{s}$ and $\SI{25}{s}$, which spans the range typically relevant to wave energy converters. However, measurement errors can still occur outside this period range, even when the conditions remain within the buoy’s operational limits.

2) The experiments reveal two error regions within the Waverider MkIII specification wave period range of $1.6$ to $\SI{30}{s}$ \citep{Datawell2025WaveriderManual}. For short periods at $T_m<\SI{5}{s}$, the $\SI{1.28}{Hz}$ sampling induces sub-Nyquist artifacts, biasing elevation and thus energy flux, and errors exceed $100\%$ for $T_m<\SI{2}{s}$. For long periods at $T_m>\SI{25}{s}$, the buoy elevation is attenuated relative to the commanded motion; simultaneous agreement of raw acceleration with the prescribed motion points to internal double-integration postprocessing as the dominant source. The wave height error can reach $\sim\!65\%$ and is largely independent of amplitude in this region.

3) Sampling rate sensitivity is modest for $\SI{5}{s}< T_m<\SI{25}{s}$: across all four methods, energy flux calculations agree within about $2\%$ between $\SI{100}{Hz}$ and $\SI{1.28}{Hz}$. In the long-period region, the sampling rate itself is not the main driver, but the elevation amplitude bias dominates. Using the small-amplitude scaling $J\propto \eta_{{rms}}^2$ and the definition of $\sigma$, the relative energy error obeys $\Xi=2\sigma-\sigma^2$. A regression over the long-period tests gave $100\sigma \approx 12.81 T_m - 321.50$ and $100\Xi \approx -1.67 T_m^2 + 109.34 T_m -1692.92$ for $T_m\!\in\![\SI{25}{s},\SI{30}{s}]$, implying that uncorrected swell can lead to large underestimation of wave energy flux $J$.

4) Wave energy flux estimation analysis shows that, when the IEC frequency-domain estimate at $\SI{100}{Hz}$ is used as the baseline, the estimates obtained at $\SI{1.28}{Hz}$ typically differ by less than $6\%$, provided that the energy is not concentrated at very short periods. The Hilbert method underestimates $J$ by $6.90\%$, $2.92\%$, and $14.40\%$ for Jennette’s Pier, WETS, and PacWave, respectively, whereas the small-amplitude and wavelet methods mildly overpredict energy flux. Practically, short-period conditions should not rely on \SI{1.28}{Hz} elevation without uncertainty inflation, while long-period conditions warrant calibration via $\sigma(T_m)$ before using wave energy flux for resource characterization.

Overall, the present study shows that the DWR-MkIII buoy exhibits two primary error branches at short and long wave periods, with the maximum energy flux error exceeding $80\%$. Moreover, the frequency-domain method recommended in the IEC standard may introduce methodological uncertainty into wave energy estimation.

The present study isolates vertical heave in a controlled environment, and real seas introduce multi-frequency and directional content and mooring dynamics. Building on these findings, future work is planned:  (1) Conduct ocean basin experiments to quantify another wave measurement uncertainty arising from hydrodynamic properties and mooring effects; (2) Investigate the high-period measurement error propagation to the frequency domain wave energy flux estimate and propose a corresponding correction method; (3) The results indicate that wave energy flux estimates differ across methods. A more detailed methodological uncertainty analysis is needed to quantify method-dependent biases and to identify an appropriate approach. The long-term goal of this research is to identify uncertainty sources that can inform the development of marine-energy-related IEC standards contributing to the marine energy development.

\section*{Acknowledgments}
This work was authored by the National Laboratory of the Rockies for the U.S. Department of Energy (DOE), operated under Contract No. DE-AC36-08GO28308. Funding provided by the U.S. Department of Energy Office of Energy Efficiency and Renewable Energy Water Power Technologies Office. The views expressed in the article do not necessarily represent the views of the DOE or the U.S. Government. The U.S. Government retains and the publisher, by accepting the article for publication, acknowledges that the U.S. Government retains a nonexclusive, paid-up, irrevocable, worldwide license to publish or reproduce the published form of this work, or allow others to do so, for U.S. Government purposes. A portion of the research was performed using computational resources sponsored by the U.S. Department of Energy’s Office of Critical Minerals and Energy Innovation and located at the National Laboratory of the Rockies.

\section*{Declaration of Competing Interests}
The authors declare no competing interests.

\appendix
\section{Test matrix iteration based on Bayesian optimization}\label{appendixgp}

\begin{figure}[ht!]
    \centering
    \includegraphics[width=.8\textwidth]{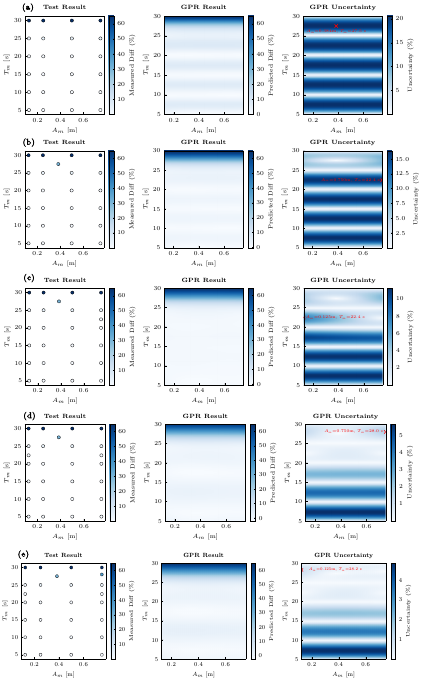}
    \caption{Bayesian optimization iteration results for test matrix selection. From left to right: scatter test results, $\mathcal{GP}$ regression result, Variance of $\mathcal{GP}$ results based on \cref{gpvirance}. (a) Maximum variance location: $A_m = \SI{0.384}{m}, T_m = \SI{27.5}{s}$; (b) maximum variance location: $A_m = \SI{0.750}{m}, T_m = \SI{22.4}{s}$; (c) maximum variance location: $A_m = \SI{0.125}{m}, T_m = \SI{22.4}{s}$; (d) maximum variance location: $A_m = \SI{0.750}{m}, T_m = \SI{28.0}{s}$; (e) maximum variance location: $A_m = \SI{0.125}{m}, T_m = \SI{28.2}{s}$.}
  \label{gpresult}
\end{figure}

Five Bayesian optimization iterations are performed using the Adaptive Computing framework \citep{griffin2025adaptive}. The acquisition function, which constitutes the selection criterion, employed is the maximum predictive variance, as defined in \cref{gpvirance}. Additionally, the constraint $T_m>\SI{20}{s}$ was also imposed to target the high-period error region:
\begin{equation} \label{finalcri}
\begin{aligned}
(T_m, A_m)_{next}
  &= \argmax_{(T_m,A_m)} \; s_*(T_m,A_m) \\
\text{s.t.}\quad & T_m > \SI{20}{s}.
\end{aligned}
\end{equation}

\cref{gpresult} presents the five sequential Bayesian optimization iterations used to choose the additional test cases. In each row, the left subfigure shows the previous tests at that iteration (circles), the middle subfigure shows the $\mathcal{GP}$ posterior mean $\mu_*(T_m,A_m)$ of the error surface $\sigma(T_m,A_m)$, and the right subfigure shows the predictive variance $s_*^2(T_m,A_m)$ based on \cref{gpvirance}. The selected subsequent experimental conditions, i.e., at the location of maximum predictive variance based on \cref{finalcri}, are indicated with a red ``x."

With the constraint $T_m>\SI{20}{s}$, the variance field forms an obvious ridge in the long-period region that is nearly insensitive to amplitude, consistent with our finding that $\sigma$ depends primarily on $T_m$ and only weakly on $A_m$. The five selected points are
$(A_m,T_m)=(\SI{0.384}{m},\SI{27.5}{s})$,
$(\SI{0.750}{m},\SI{22.4}{s})$,
$(\SI{0.125}{m},\SI{22.4}{s})$,
$(\SI{0.750}{m},\SI{28.0}{s})$,
and $(\SI{0.125}{m},\SI{28.2}{s})$.

Incorporating each new observation progressively collapses the long-period uncertainty ridge while preserving the weak sensitivity to amplitude, thereby efficiently refining $\sigma(T_m,A_m)$ where errors are most consequential. This method will be applied in future multi-DOF experiments to improve the experimental efficiency.


\end{document}